# Ekman Theory with Damping

**Authors**

Jiacheng Wu,[1,2] Yonggang Liu,[1,2*] Ruixin Huang,[3] Jinhan Xie,[4] Zhaoying Wang,[5] Shaoqing Zhang,[5*]

**Affiliations**

[1]Laboratory for Climate and Ocean-Atmosphere Studies, Department of Atmospheric and Oceanic Sciences, Peking University; Beijing, 100871, China.

[2]Institute of Ocean Research, Peking University, Beijing, 100871, China.

[3]Woods Hole Oceanographic Institution, Woods Hole, MA, USA

[4]Department of Mechanics and Engineering Science, College of Engineering, Peking University, Beijing, 100871, China.

[5]Key Laboratory of Physical Oceanography, The College of Oceanic and Atmospheric Sciences, MOE, Institute for Advanced Ocean Study, Frontiers Science Center for Deep Ocean Multispheres and Earth System (FDOMES), Ocean University of China, Qingdao, China

*Corresponding author: Yonggang Liu and Shaoqing Zhang

**Email:** ygliu@pku.edu.cn; szhang@ouc.edu.cn;

**Abstract**

The observed Ekman spirals in the ocean are always "flatter" than that predicted by the classic theory. We propose that the universal flattening of Ekman spiral is mainly due to the damping associated with turbulent dissipation. Analytical solutions and numerical simulations show convincingly a better fitting between the new theory and observations. Most importantly, the new theory indicates that the damping can lead to weakened Ekman transport and pumping, with the latter not only driven by the curl but also the divergence of wind stress. Under a modest damping, the Ekman transport along 26.5°N will be ~0.4 Sv (12%) smaller than that predicted by the classic theory. Hence, the damping due to turbulent dissipation can noticeably affect the wind-driven circulation in the upper ocean.

# MAIN TEXT

## Introduction

The classic Ekman theory predicts that the surface ocean current deviates from the wind stress by 45° (to the right in the Northern Hemisphere)[1]. The theory also predicts that the speed of ocean current decays exponentially and the current direction rotates continuously with depth (known as Ekman spiral), so that the vertically integrated ocean transport (Ekman transport) is perpendicular to the surface wind. The divergence of Ekman transport induces a vertical velocity known as Ekman pumping. The Ekman transport and pumping largely explain the wind-driven circulation in the upper ocean, including the gyre circulation and upwelling/downwelling systems. The Ekman theory is thus critically important for understanding the ocean dynamics, marine ecosystem, and climate.

Despite the power of the Ekman theory in explaining qualitatively many of the phenomena in the ocean, quantitative verification of the Ekman spiral from oceanic observations proved challenging[2-6]. One of the most significant discrepancies between theory and observations is that the observed speed of Ekman current decreases with depth at a rate that is always faster than that of the current rotation[3]. This phenomenon has been termed as flattening or compression of

Ekman spiral[4, 7-9]. Here we propose that this issue is due to the underestimation of turbulent damping in spite of the detailed causes of the damping, in the classic Ekman theory.

The classic Ekman theory assumed that the surface ocean is at steady state and horizontally homogeneous, and the ocean properties like density and viscosity are constant with depth[1]. The theory also ignored the influence of waves and their associated Stokes current. Previous studies have tried to improve the model-observation comparison by considering more complexities such as vertically varying viscosity[10-16], the anisotropy of viscosity[17], the unsteadiness of wind[13, 18], the Stokes current and Langmüir circulation[19-21], the geostrophic shear[8, 22, 23], and the stratification [9]. These modifications all affect the structure of the Ekman spiral and may cause flattening of the spiral.

The complexities added above are more realistic than the classic theory, but are often specific to certain observation or unconstrained and may even be inconsistent with each other. For example, some studies used linearly increasing viscosity with depth[10, 11], while others used linearly or nonlinearly decreasing viscosity[13, 22]. Some studies thought that the non-constant viscosity is unnecessary when compared to observations[8, 18, 24]. Some study invoked the time-varying stratification as the key to understanding the flattening of the Ekman spiral[9], while others found no clear relationship between spiral structure and stratification[8]. Moreover, no single study could include all the complexities mentioned above appropriately because there would be too many parameters to be constrained by observations; the effect of the chosen complexities was also often overstated (see Fig. 3 of [11] and Figs. 3b and 4b of [24]). Furthermore, simple analytical solutions that can give immediate insight to the consequence of the additional complexities described above are very difficult to obtain, even though many of them are quite elegant already[11, 18, 23]. The exceptions are the Stokes current and geostrophic shear, the influence of which over most part of the water column have been shown to be equivalent to a modification of the surface wind stress[20, 22].

Since the flattening of Ekman spiral is a universal phenomenon in the real ocean[3-5, 7, 8], there must be a process that is acting independent of the local complexities mentioned above, or the different local processes may have the same effect and share the same expression in the Ekman equation. As is demonstrated in Materials and Methods and SI Appendix, the flattening of Ekman spirals is most easily explained by considering additional linear damping in the equation. This damping could come from lateral viscous friction or other processes that have been considered in previous studies (e.g., non-constant vertical viscosity). We will demonstrate that this damping naturally causes flattening of Ekman spiral and the solution of this new Ekman theory can fit the observed Ekman spiral reasonably well at many sites representative of all typical situations of the ocean. Moreover, the solutions indicate that the Ekman transport has been overestimated in the classic theory and, in addition to wind stress curl, the Ekman pumping should also depend on the divergence of wind stress.

## Results
### Ekman spiral with damping

The universally existing flattening of the Ekman spirals in the real ocean suggests that there was an important force neglected in the classic Ekman equation, and this force is best represented by a linear damping term (see the derivation and reasoning in Materials and Methods and SI Appendix). Introduce the linear damping term into the basic equation (see SI Appendix) describing the Ekman spiral,

$$ifU = A_z \frac{\partial^2 U}{\partial z^2} - RU \qquad (1)$$

where $U = u + iv$ is the velocity in complex form, $i$ is the imaginary number, $f$ is the Coriolis parameter, $A_z$ is the vertical eddy viscosity, and $R$ is the damping coefficient. As shown in the SI Appendix, subjected to the same boundary conditions as in the classic theory, the solution is,

$$U = \mu U_{Ek} \exp\left\{\frac{z}{d_{amp}} + i\left[\text{sgn}(f)\left(\frac{z}{d_{rot}} - \theta\right) + \phi_\tau\right]\right\} \qquad (2)$$

where $U_{Ek} = d_{Ek}\frac{|\tau|}{\sqrt{2}\rho A_z}$ is the amplitude of surface current in the classic Ekman theory with $d_{Ek} = \sqrt{\frac{2A_z}{|f|}}$ the depth scale of the classic Ekman theory, $\tau$ the surface wind stress, and $\rho$ the density of seawater. $\mu = (1 + Da^2)^{-\frac{1}{4}}$ is a factor by which the surface velocity is reduced due to damping (see detailed derivation in SI Appendix and its value from Fig. 1D), in which $Da = \frac{R}{|f|}$ is a non-dimensional number measuring the ratio between timescales of inertial motion and damping. $\theta = \frac{1}{2}\arctan Da^{-1}$ is the surface deflection angle (Fig. 1C), clockwise from the wind stress whose direction $\phi_\tau$ is measured counterclockwise from east. Since $Da > 0$, the new Ekman velocity in Eq. 2 is always smaller than the classic one, and the surface deflection angle is smaller than 45°. It can be seen from Eq. 2 that the key parameter is $Da$, and is defined as the damping number hereafter.

In the classic Ekman theory there is one depth scale $d_{Ek}$ only; however, for the case with damping, there are two depth scales, $d_{amp}$ for the amplitude and $d_{rot}$ for the rotation, with expression,

$$d_{amp} = \mu d_{Ek}\frac{\cos\frac{\pi}{4}}{\cos\theta}, \quad d_{rot} = \mu d_{Ek}\frac{\sin\frac{\pi}{4}}{\sin\theta} \qquad (3)$$

Since $\mu<1$ and $\theta<45°$, $d_{amp}$ is smaller than $d_{Ek}$, while $d_{rot}$ is larger than $d_{Ek}$ (Fig. 1), with $d_{amp}d_{rot} = d_{Ek}^2$. Therefore, the flattening of Ekman spiral as observed in the ocean is a natural result of the damping. When the damping number is 0, the variables in both Eqs. 2 and 3 recover to those in the classic Ekman theory.

The Ekman transport and Ekman pumping with damping can be obtained straightforwardly,

$$T = T_x + iT_y = \mu^2 T_{Ek}\exp\left[i(\phi_\tau - \text{sgn}(f)2\theta)\right] \qquad (4)$$

$$w = \mu^2\left[\nabla_z\cdot\left(\frac{\tau}{\rho|f|}\right)\cos 2\theta + w_{Ek}\sin 2\theta\right] \qquad (5)$$

where $T_{Ek} = \frac{|\tau|}{\rho|f|}$ and $w_{Ek} = \text{sgn}(f)\nabla_z\times\left(\frac{\tau}{\rho|f|}\right)$ are the magnitude of the classic wind driven Ekman transport and pumping, the subscript $z$ in $\nabla_z$ means the operation is for the horizontal directions only. Thus, both the Ekman transport and pumping are smaller than the classic solutions by a factor $\mu^2$. Note that $T$ is now not perpendicular to wind stress. Importantly, the Ekman pumping is now not only dependent on the wind stress curl, but also the divergence of wind stress.

**Comparison with the observed Ekman spiral**

The damping-enabled Ekman theory developed here allows us to inverse the depth scales $d_{amp}$ and $d_{rot}$ (fig. S1), as well as $A_z$ and $R$, from the observations of $u(z)$ and $v(z)$ (Materials and Methods and SI Appendix). The wind stress in here is determined by substituting the observed velocity at a certain depth into Eq. 2 rather than from the observations directly because 1) the observed wind stress may be subject to large uncertainties and 2) the effect of Stokes flow may cause the whole spiral to be rotated by a certain angle relative to that induced by the surface wind[20]. Observed Ekman spirals at five locations are used (table S1), they are at the south of the gulf stream[3], western Pacific near the equator[7], the coast of northern California[4], Drake Passage[5], and north of Kerguelen[8]. The observations thus cover different oceanic systems: eastern (site A in Fig. 2) and western (site B) boundary current, equatorial current (site C), the Antarctic Circumpolar Current (ACC; sites D and E). All parameters used and their values are shown in table S2.

Fig. 2 shows that the structure of Ekman spirals predicted by the new theory fits the observed data points well, much better than that by the classic theory. The quality of comparison at near Gulf Stream and coast of California (Fig. 2A,B) are visibly superior than that in [11] who tried to include the influence of unsteadiness of winds, linear variation of vertical viscosity with depth, and the Stokes current. The comparison is also not worse than that in [9] who considered the influence of the variation of both vertical viscosity and stratification. The comparison at Drake Passage (Fig. 2D) is again better than that in [24] who tried to improve the fit with the classic Ekman theory by correcting the influence of geostrophic shear. A good fit is not obtained near the equatorial Pacific (Fig. 2C); the fit was not improved by considering unsteadiness of winds and oceanic stratification, vertical variation of vertical viscosity, and Stokes current either[9, 11], with no clear reasons.

**Flattening of Ekman spiral and damping in numerical simulations**

The theory-predicted Ekman spiral is also compared to those obtained in both an idealized numerical experiment conducted using an oceanic general circulation model (GCM), MITgcm (Figs. S2 and S3; Materials and Methods and SI Appendix) and a realistic eddy-permitting global simulation performed using an atmosphere-ocean GCM, CESM1.3 (Materials and Methods and SI Appendix). In the idealized experiment, the situation is simplified by specifying steady wind and constant density and vertical viscosity. Moreover, there are no waves, so that the near surface currents are not complicated by the presence of Stokes current. Under such idealized situation, flattening of Ekman spiral is also observed. The new theory produces a much better fit than the classic theory (fig. S3), utilizing $R$ estimated simply by $R = \frac{1}{2}\left(\frac{A_h \nabla^2 u}{u} + \frac{A_h \nabla^2 v}{v}\right)$, where $u$ and $v$ are the zonal and meridional velocities over the first 10 m, and $A_h$ is the horizontal mixing coefficient used in the simulation. That is, no inversion using SI Appendix Eqs. S17-S18 is even needed. The nice fit here means that the additional damping (second term of the righthand side of Eq. (1)) in this idealized case is primarily caused by lateral viscous friction.

In the realistic global simulation, the flattening of Ekman spirals at the same locations as those of the observed ones described above can be analyzed. Strong flattening is found at all five locations (fig. S4). Once again, the new theory produces a much better fit than the classic theory. In this case, the parameter $Da$ is determined in the same way as that for the observations, but the values obtained (table S3) are different from those in the observations (table S2). This is understandable since the global simulation here is not an exact reproduction of the observations. Other than lateral friction, the additional damping in this realistic case may also come from temporal variations of vertical stratification and viscosity, geostrophic shear, anisotropic viscosity etc. All these processes have been mentioned in the literature in different regions.

The global simulation allows us to explore whether the parameter $R$ or $Da$ is related to the damping in the model, which was not possible for the observations. The damping in the model can be estimated using the second-order longitudinal structure function[25], from which a parameter $\xi$ can be obtained for each location (fig. S5; Materials and Methods and SI Appendix). This parameter $\xi$ is proportional to the two-thirds power of the energy dissipation rate (Eq. S20), thus having different meaning from $R$. Inferring $R$ from $\xi$ is not possible because the proportion coefficient in the Eq. S20 depends on the dissipation rate itself, and even if the dissipation rate is obtained from $\xi$, the low temporal resolution (monthly) of the velocity data does not allow us to obtain $R$. Therefore, we can only compare $\xi$ and $R$ qualitatively. Because $\xi$ and $R$ (or $Da$) are estimated completely independently, it will provide some support to our theory if they are proportional to each other. Indeed, Fig. 5B shows that $\xi$ estimated from the model surface velocity and $Da$ estimated from the modeled Ekman spirals are proportional. The relationship between $\xi$ and $Da$ seems better than that between $\xi$ and $R$ (Fig. 5A), probably indicating that the damping in the ocean is related to the Coriolis parameter $f$. Nevertheless, the good relationship between $\xi$ and $Da$ may serve as evidence that the damping is a significant contributor to the flattening of Ekman spirals.

**Implication: Ekman transport and Ekman upwelling**

The classic Ekman theory is often used to diagnose the meridional transport (e.g. RAPID Array; https://rapid.ac.uk) and upwelling of the oceans [e.g. 26]. The damping-enabled Ekman theory shows that the Ekman transport should be smaller than that predicted by the classic Ekman theory (Eq. 4), and the magnitude of upwelling should be smaller in general too unless the influence of the wind-stress divergence is large and reinforcing (Eq. 5). If the damping-enabled Ekman theory is valid, then the possible influence of damping on Ekman transport and pumping in the real ocean can be estimated for a series of different $Da$.

The zonally integrated Ekman transport along different latitudes in the Southern Ocean is estimated and shown in Fig. 3A. The maximum value averaged over 1991-2020 is 42.81 Sv, around 48°S. This value will be 2.62 Sv smaller if $Da$ is 0.2, and 7.65 Sv smaller if $Da$ is 0.4. The time series of zonally integrated Ekman transport along 26.5°N, where RAPID array is installed, is also calculated for $Da$ is 0.2. This transport is often used in estimating the magnitude of Atlantic meridional overturning circulation [AMOC27]; at this latitude. Fig. 3B, C shows that the annual mean of this transport will be overestimated by ~0.4 Sv (12%) overall, with the maximum overestimation reaching 0.5 Sv.

The influence of damping on Ekman pumping is also significant, with the largest correction obtained near the equator (Fig. 4C); when $Da$ is 0.2, the correction can be as large as ±0.1 m/day (Fig. 4C). In relatively terms, the correction can be greater than 80% (Fig. 4D). The positive values in Fig. 4D indicate that although the damping induces a weakening in general, it is possible that the pumping be strengthened. This strengthening is due to the influence of wind stress divergence (Eq. (5)) which does not exist in the classic Ekman theory. Therefore, the influence of damping on Ekman transport and damping could be an important correction to consider even if the damping number $Da$ is not large.

**Discussion**

The value for $R$ near the equatorial Pacific (site C in Fig. 2) is found to be ~$7 \times 10^{-6}$ s$^{-1}$, giving a damping timescale ($1/R$) of 39.7 hours. This timescale denotes the time it takes for a body of unforced moving water to quiet down, and thus may be compared to the decay timescale of the inertial oscillation. The smaller value of $R$ found here has the same order of magnitude as that used in McWilliams and Huckle [13] but they were not focusing on the flattening of Ekman spiral. According to the solution here (Eq. 2), the smaller surface deflection angle (31° and 33°) obtained in the numerical simulation in [13] was at least partially due to damping since they explicitly included this effect in their equations. The corresponding damping number $Da$ at site C is 0.28 (table S2), which implies a moderate correction to the Ekman transport and upwelling (Fig. 4). However, $Da$ is greater than 0.5 at all other points (table S2). The large $Da$ at these sites may be due to two reasons: 1) Ekman flow has larger vertical shear than the inertial current, and hence experiences stronger damping, 2) these sites are close to a coast (Fig. 2) which greatly enhances damping as shown by the decay of mesoscale eddies [e.g.28].

The flattened Ekman spiral indicates that the force due to vertical shear is not aligned with the Coriolis force for any sub-layer of water within the Ekman layer. Such phenomenon has also been interpreted as the effect of a complex viscosity, which will rotate the shear stress by an angle in order for the two forces above to remain aligned with each other[5, 9, 21]. A complex viscosity is physically not meaningful, and thus not liked [e.g. 9]. Such awkwardness can be removed by showing that the complex viscosity is essentially the same as adding a damping term in the momentum equation (SI Appendix). Therefore, the damping-enabled Ekman theory introduced herein provides a physically more straightforward way of explaining the flattening of Ekman spiral.

The success of the theory (Eq. (1)) in explaining the observed and simulated flattening of Ekman spiral may indicate that the major effect of various processes that have been proposed in the literature can be lumped into a linear damping. The lateral dissipation, have not been mentioned yet, should also be included. These processes may cause extra damping, more than already considered by the classic theory (the first term of the right-hand side of Eq. (1)). In the meantime, these processes may also other effect and distort the Ekman spirals differently (there are still mismatch between the observed and theory-predicted Ekman spirals). Currently, it is difficult to determine the value of $R$ or $Da$, and thus a new unknown is introduced to the Ekman theory. Nevertheless, the universal flattening of Ekman spirals is well explained by a simple rectification to the classic Ekman theory, and important implications of the additional damping can be derived from the new theory. Moreover, the value of the unknown parameter may be determined in the future when high-resolution observations of surface velocity in both time and space become available.

**Materials and Methods**

**The implication of flattened Ekman spirals and the damped Ekman model.**

The observed Ekman spirals indicate that the depth scales for the decay of amplitude $d_{amp}$ and the rotation of velocity vector $d_{rot}$ (see SI Appendix Eq. S1) are not equal. Moreover, Lenn and Chereskin [5] showed that the observed vertical eddy viscosity $A_z$ should be a complex value which mean that there should be an additional force to balance the vertical shear stress and Coriolis force. We mathematically prove that the mismatch between depth scales $d_{amp}$ and $d_{rot}$ is equivalent to simply adding a linear damping term in the classic Ekman model. The full derivation of the damped Ekman model and how we inverse the spiral from the observations are provided in SI Appendix.

**Model and configuration of the idealized experiment and the realistic global experiment.**

The oceanic general circulation model MITgcm (https://mitgcm.readthedocs.io/en/latest/) is used to study the possible flattening of Ekman spiral. In our idealized experiment carried out herein, the advection is turned off in the momentum equation to be consistent with the Ekman model. The experimental domain is rectangular. *f*-plane approximation is used with $f = 10^{-4}$ s$^{-1}$, equivalent to a latitude of 43°N. The water density is 1000 kg m$^{-3}$ and does not change with time and space. No-slip condition is specified at lateral boundaries; at the surface, zonal (i.e., along *x* direction) wind stress is specified which has a value of 0.1 N m$^{-2}$ at y = 0 and decreases to 0 at the northern and southern boundaries according to a cosine function. To see whether flattening of Ekman spirals in realistic simulations is comparable to that in the observations and whether the damping is indeed an important cause of the flattening, we utilize the results from a high-resolution simulation[29] with the Community Earth System model version 1.3 (CESM1.3). The simulation was done with a horizontal resolution of approximately 0.25° for the atmosphere and land and a nominal resolution of 0.1° for the ocean and sea-ice. The ocean component has 62 levels in the vertical with a maximum depth of 5875 m and the spacing is 10 m in the upper 100 m. The horizontal resolution is high enough so that the damping can be estimated using the second-order longitudinal structure function (see next section) and the vertical resolution is sufficient for demonstrating the Ekman spirals. The ocean wave model is stub only in the CESM1.3. The simulation is a "free-run" without any data assimilation and covers the period from 1850 to 2100, using time-varying historical greenhouse gas and anthropogenic and natural aerosol forcing from 1850 to 2005 and representative concentration pathway 8.5 (RCP8.5, a high greenhouse gas emission scenario) forcing during 2006–2100. Both simulations results flattened Ekman spiral which can be fitted well in the Ekman-damping model. Further details on the model simulations are provided in SI Appendix.

**Diagnosing damping in the realistic global experiment**

It is essential to demonstrate that the damping in the model simulations is well linked to the damping parameter *R* or *Da* inferred directly from the simulated flatten Ekman spirals. However, because only monthly data are available from the realistic global simulation, the damping has to be estimated indirectly from the kinetic energy dissipation rate. For this purpose, the second-order longitudinal structure function (see SI Appendix Eq. S19[25]) is a powerful tool (see SI Appendix for more detail)

**References**
1. Ekman VW. On the influence of the earth's rotation on ocean-currents. 1905.
2. Stacey MW, Pond S, LeBlond PH. A wind-forced Ekman spiral as a good statistical fit to low-frequency currents in a coastal strait. *Science*. 1986; **233**(4762): 470-472.
3. Price JF, Weller RA, Schudlich RR. Wind-driven ocean currents and Ekman transport. *Science*. 1987; **238**(4833): 1534-1538.
4. Chereskin T. Direct evidence for an Ekman balance in the California Current. *Journal of Geophysical Research: Oceans*. 1995; **100**(C9): 18261-18269.
5. Lenn Y-D, Chereskin TK. Observations of Ekman currents in the Southern Ocean. *Journal of Physical Oceanography*. 2009; **39**(3): 768-779.
6. Schudlich RR, Price JF. Observations of seasonal variation in the Ekman layer. *Journal of physical oceanography*. 1998; **28**(6): 1187-1204.
7. Wijffels S, Firing E, Bryden H. Direct observations of the Ekman balance at 10 N in the Pacific. *Journal of Physical Oceanography*. 1994; **24**(7): 1666-1679.
8. Roach CJ, Phillips HE, Bindoff NL *et al*. Detecting and characterizing Ekman currents in the Southern Ocean. *Journal of Physical Oceanography*. 2015; **45**(5): 1205-1223.
9. Price JF, Sundermeyer MA. Stratified ekman layers. *Journal of Geophysical Research: Oceans*. 1999; **104**(C9): 20467-20494.

10. Madsen OS. A realistic model of the wind-induced Ekman boundary layer. *Journal of Physical Oceanography*. 1977; **7**(2): 248-255.
11. Lewis D, Belcher S. Time-dependent, coupled, Ekman boundary layer solutions incorporating Stokes drift. *Dynamics of atmospheres and oceans*. 2004; **37**(4): 313-351.
12. Zikanov O, Slinn DN, Dhanak MR. Large-eddy simulations of the wind-induced turbulent Ekman layer. *Journal of Fluid Mechanics*. 2003; **495**: 343-368.
13. McWilliams JC, Huckle E. Ekman layer rectification. *Journal of physical oceanography*. 2006; **36**(8): 1646-1659.
14. Constantin A, Dritschel DG, Paldor N. The deflection angle between a wind-forced surface current and the overlying wind in an ocean with vertically varying eddy viscosity. *Physics of Fluids*. 2020; **32**(11).
15. Shrira VI, Almelah RB. Upper-ocean Ekman current dynamics: a new perspective. *Journal of Fluid Mechanics*. 2020; **887**: A24.
16. Dritschel DG, Paldor N, Constantin A. The Ekman spiral for piecewise-uniform viscosity. *Ocean Science*. 2020; **16**(5): 1089-1093.
17. Huang RX. *Ocean circulation: wind-driven and thermohaline processes*: Cambridge University Press, 2010.
18. Elipot S, Gille S. Ekman layers in the Southern Ocean: Spectral models and observations, vertical viscosity and boundary layer depth. *Ocean Science*. 2009; **5**(2): 115-139.
19. Huang NE. On surface drift currents in the ocean. *Journal of Fluid Mechanics*. 1979; **91**(1): 191-208.
20. Polton JA, Lewis DM, Belcher SE. The role of wave-induced Coriolis–Stokes forcing on the wind-driven mixed layer. *Journal of Physical Oceanography*. 2005; **35**(4): 444-457.
21. McWilliams JC, Huckle E, Liang J-H *et al.* The wavy Ekman layer: Langmuir circulations, breaking waves, and Reynolds stress. *Journal of Physical Oceanography*. 2012; **42**(11): 1793-1816.
22. Cronin MF, Kessler WS. Near-surface shear flow in the tropical Pacific cold tongue front. *Journal of Physical Oceanography*. 2009; **39**(5): 1200-1215.
23. Wenegrat JO, McPhaden MJ. Wind, waves, and fronts: Frictional effects in a generalized Ekman model. *Journal of Physical Oceanography*. 2016; **46**(2): 371-394.
24. Polton JA, Lenn Y-D, Elipot S *et al.* Can drake passage observations match Ekman's classic theory? *Journal of physical oceanography*. 2013; **43**(8): 1733-1740.
25. Ni R, Xia K-Q. Kolmogorov constants for the second-order structure function and the energy spectrum. *Physical Review E—Statistical, Nonlinear, and Soft Matter Physics*. 2013; **87**(2): 023002.
26. Almeida L, Mazloff MR, Mata MM. The impact of Southern Ocean Ekman pumping, heat and freshwater flux variability on intermediate and mode water export in CMIP models: Present and future scenarios. *Journal of Geophysical Research: Oceans*. 2021; **126**(6): e2021JC017173.
27. Hirschi J, Baehr J, Marotzke J *et al.* A monitoring design for the Atlantic meridional overturning circulation. *Geophysical Research Letters*. 2003; **30**(7).
28. Evans DG, Frajka-Williams E, Naveira Garabato AC. Dissipation of mesoscale eddies at a western boundary via a direct energy cascade. *Scientific Reports*. 2022; **12**(1): 887.
29. Chang P, Zhang S, Danabasoglu G *et al.* An unprecedented set of high-resolution earth system simulations for understanding multiscale interactions in climate variability and change. *Journal of Advances in Modeling Earth Systems*. 2020; **12**(12): e2020MS002298.

**Acknowledgments**
We are grateful for the discussion with Qing Li on Langmüir circulation.

**Funding:** This work is supported by NSFC 42225606 and 423B2601.

**Author contributions:** The key idea and mathematical solutions were proposed by JW. YL, JW, and RXH wrote the original draft together. JX instructed on how to estimate the dissipation rate using turbulence theory. SZ and ZW provided the high-resolution global simulation data.

**Competing interests:** The authors declare no competing interest.

**Data and materials availability:** The GODAS data can be downloaded from https://psl.noaa.gov/data/gridded/data.godas.html. The ERA5 data can be downloaded from https://cds.climate.copernicus.eu/cdsapp#!/dataset/reanalysis-era5-single-levels?tab=overview. All data are provided with this paper.

**Figures and Tables**

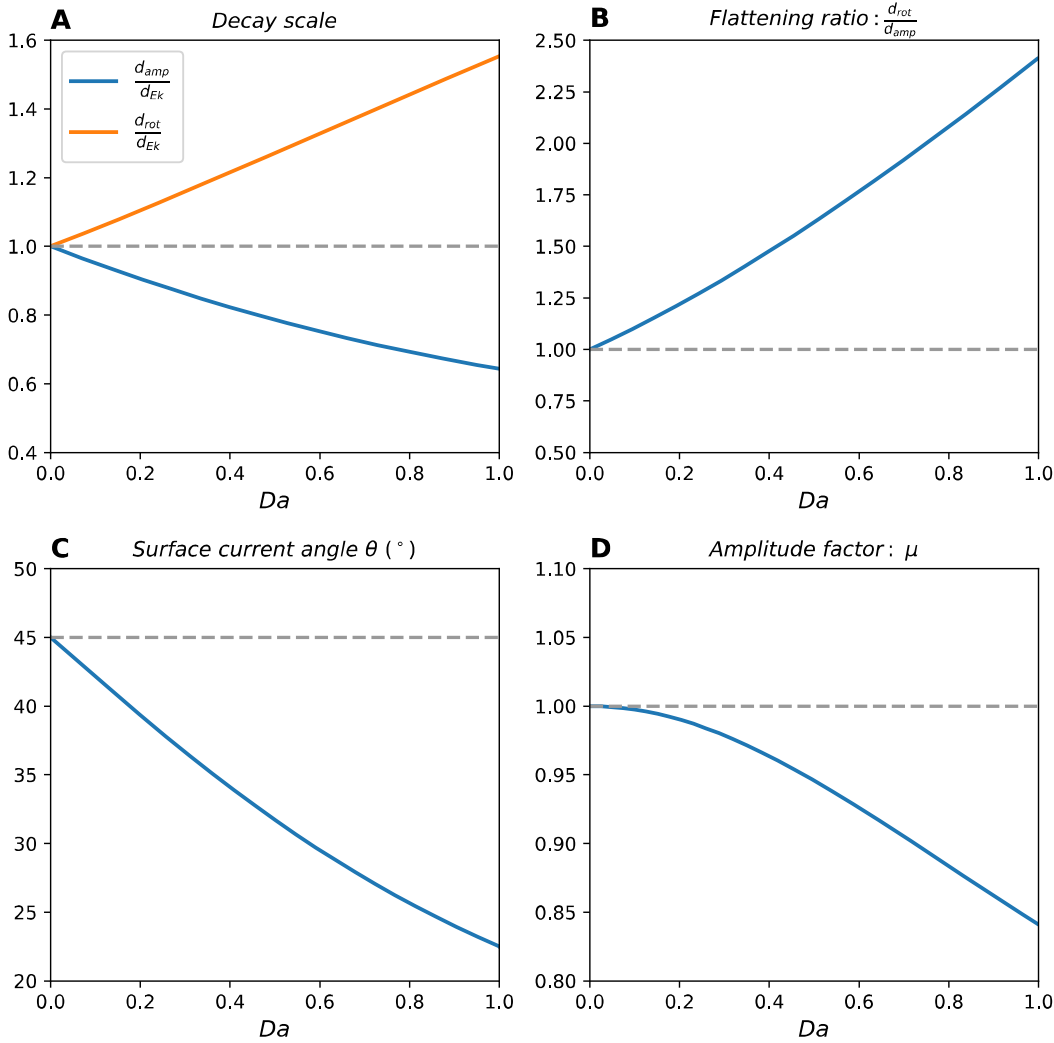

**Fig. 1. Dependence of various variables on the damping number *Da*.** (A) decay depth scales *d_amp* and *d_rot* normalized by the classic decay scale *d_Ek*. (B) flattening ratio *d_amp*/*d_rot*. (C) deflection angle of surface current. (D) the amplitude factor *μ* (solid line) in equation (2).

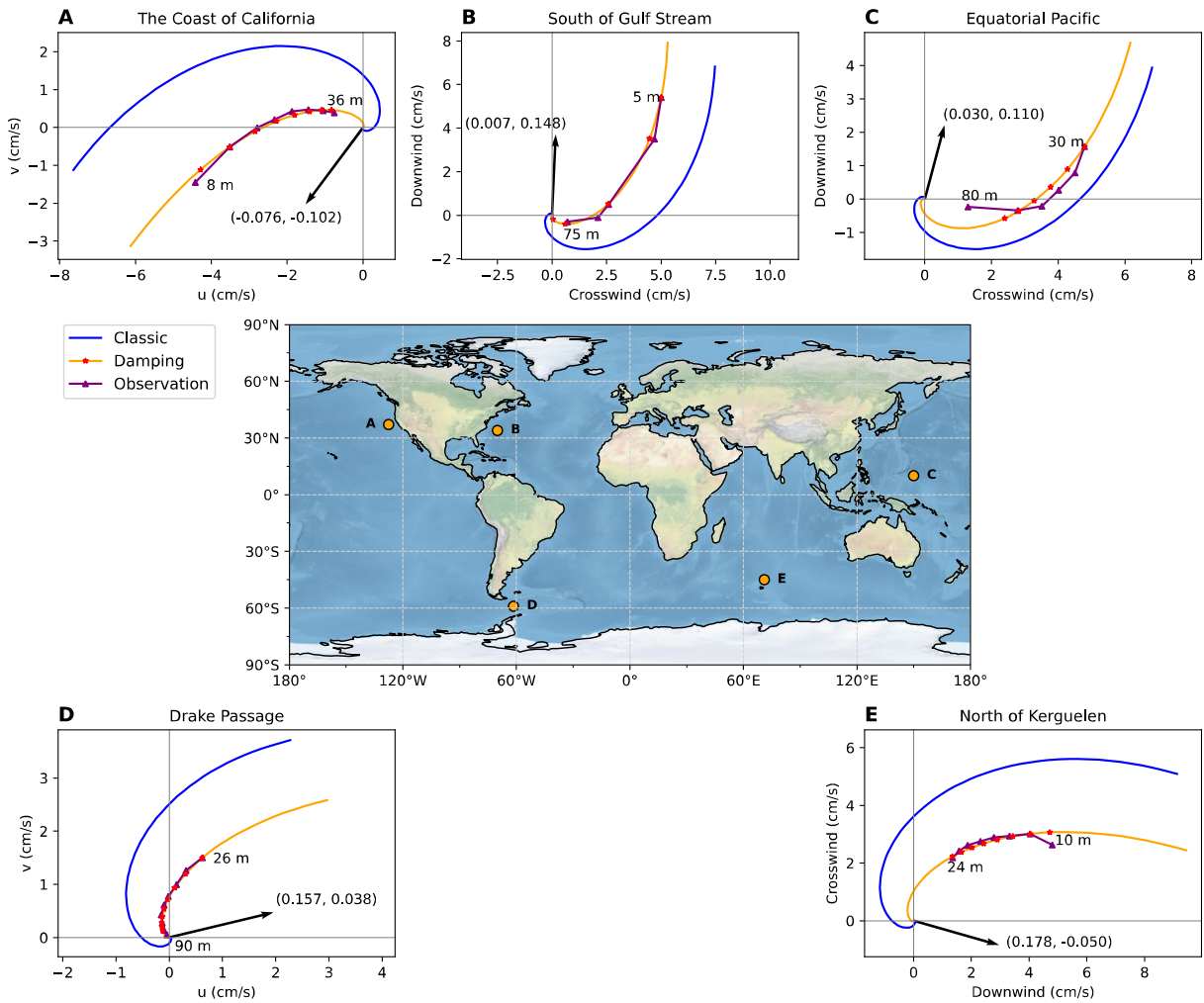

**Fig. 2. Comparison between the observed Ekman spirals (purple triangles) and the theoretical ones (blue: classic Ekman theory; orange: damping-enabled theory).** The red stars show the values calculated with the new theory at the corresponding depths of measurements. Note that the vertical resolution of each observation is different. Moreover, the x-axis and y-axis are along east and north, respectively in (A) and (D) but along the downwind and crosswind directions of the observed winds at the respective sites in (B), (C) and (E). The wind stress is shown by the black quiver with its x- and y-components given in the brackets with unit of N m-2.

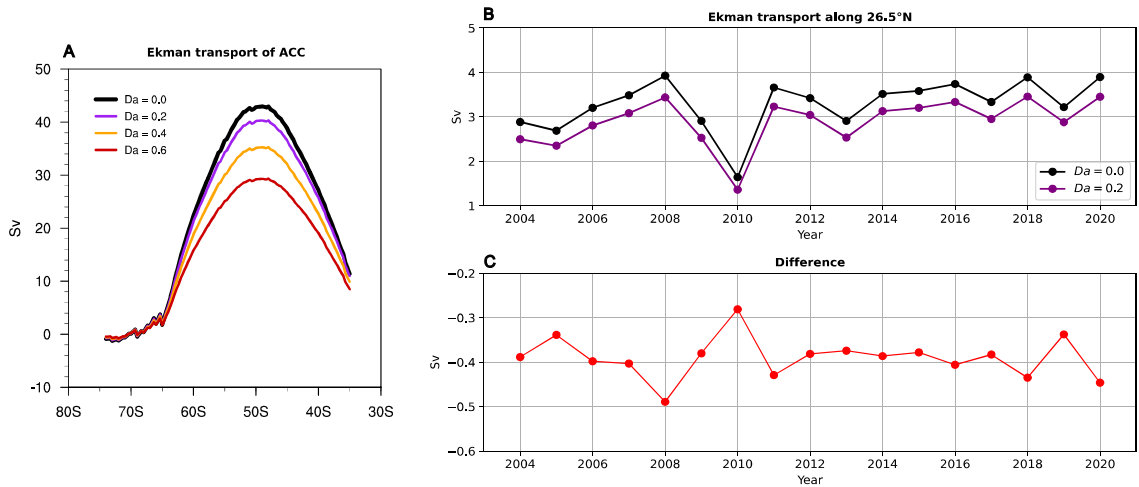

**Fig. 3. Influence of damping on Ekman transport.** (A) zonally integrated Ekman transport over the ACC region for different damping number *Da*, averaged over the period 1991-2020; The wind stress used in the calculations is from GODAS  (B) zonally integrated annual mean Ekman transport along 26.5°N, where RAPID array is installed, calculated using the classic Ekman theory (black) and damping-enabled theory (purple) with *Da* = 0.2; (C) shows the difference (purple minus black) between the two lines in (B). The wind stress used in the calculations is from ERA5

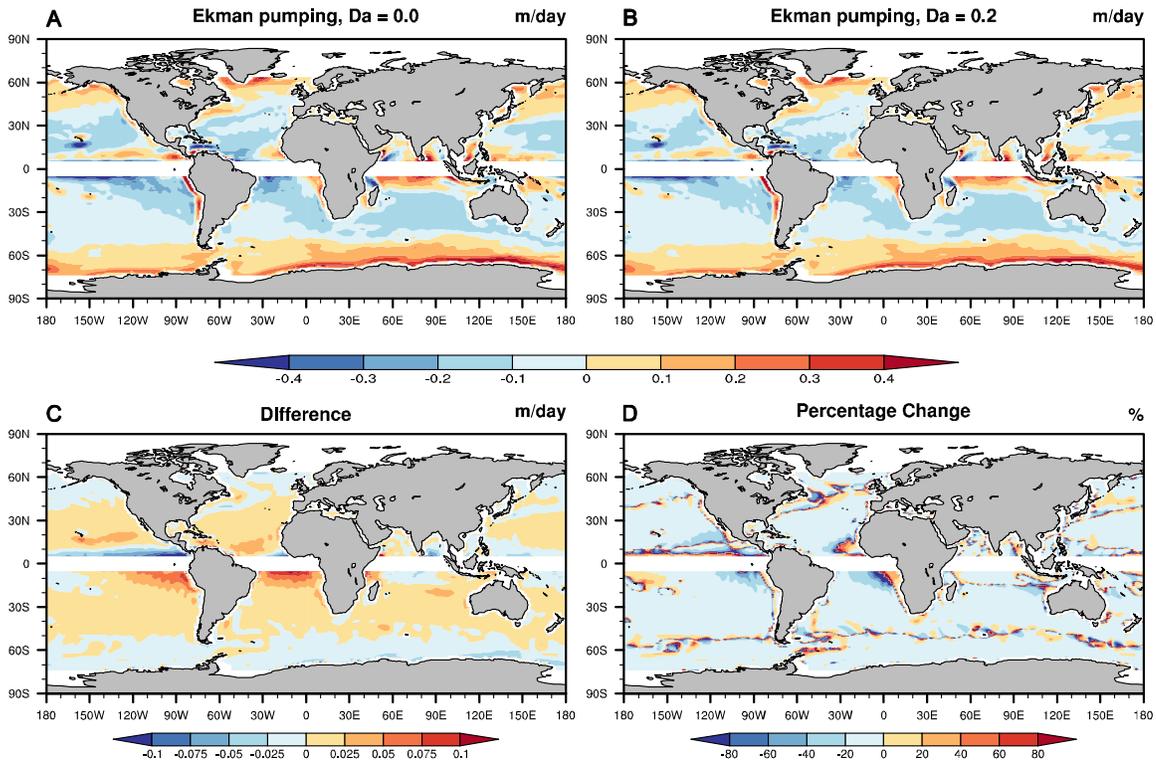

**Fig. 4. Influence of damping on Ekman pumping.** (A) and (B) show the Ekman pumping (positive upward) calculated with *Da* being 0.0 (classic Ekman theory) and 0.2, respectively. (C) shows the difference between (B) and (A) (former minus latter). (D) is the same as (C) except that here the relative changes are shown. The wind stress used in the calculations is from GODAS.

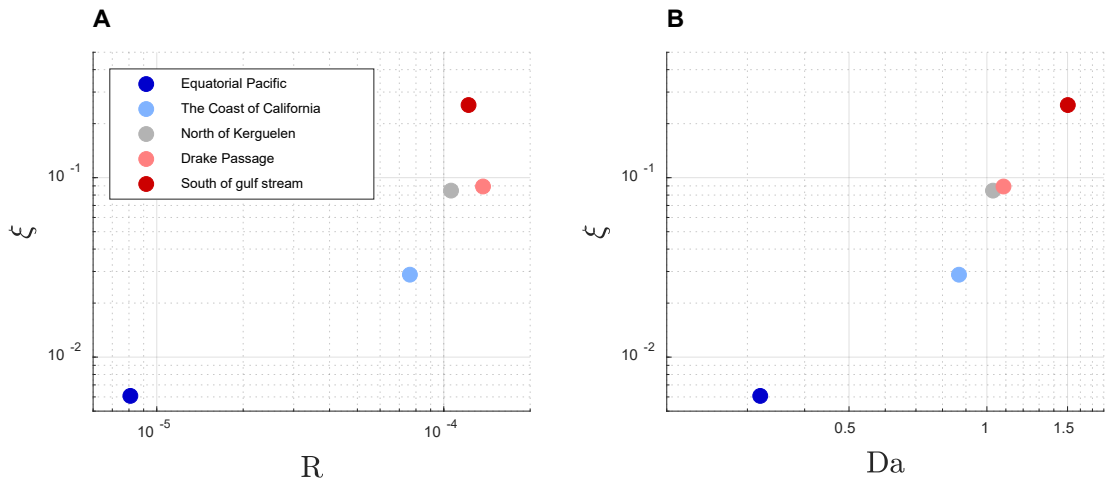

**Fig. 5. The relationship between the damping ($\xi$) estimated utilizing turbulence theory and (A) the damping coefficient $R$ or (B) the damping number $Da$.** The meaning of $\xi$ can be found from Materials and Methods and SI Appendix.

# Supplementary Materials for
## Ekman Theory with Damping

Jiacheng Wu *et al.*

*Corresponding author. Email: ygliu@pku.edu.cn; szhang@ouc.edu.cn;

**This PDF file includes:**

    Supplementary Text
    Figs. S1 to S5
    Tables S1 to S3
    References

**Supplementary Text**

**The implication of flattened Ekman spirals.**

The observed Ekman spirals indicate that the wind-induced steady state Ekman velocity varies with depth ($z$; positive upward) according to

$$u + iv = U_s \exp\left[\frac{z}{d_{amp}} + i\left(\text{sgn}(f)\frac{z}{d_{rot}} + \phi\right)\right] \quad (S1)$$

with $d_{amp}$ and $d_{rot}$ ($d_{amp} < d_{rot}$) the depth scales for the decay of amplitude and the rotation of velocity vector; $f$ is the Coriolis parameter, $U_s$ is the observed surface speed, and $\phi$ is the deflection angle between the surface velocity and wind stress. Because the observed $d_{amp}$ and $d_{rot}$ are not equal, Eq. S1 suggests that the shear stress and the velocity shear are not in parallel. This further suggests that there is an additional force missing in the classic Ekman equation (without invoking complexities such as non-constant viscosity),

$$\begin{cases} -fv = A_z \dfrac{\partial^2 u}{\partial z^2} + F_u \\ fu = A_z \dfrac{\partial^2 v}{\partial z^2} + F_v \end{cases} \quad (S2)$$

where $A_z$ is the vertical eddy viscosity. Substituting Eq. S1 into Eq. S2 gives

$$\begin{cases} \left(-f + \text{sgn}(f)\dfrac{2A_z}{d_{amp}d_{rot}}\right)v = A_z\left(\dfrac{1}{d_{amp}^2} - \dfrac{1}{d_{rot}^2}\right)u + F_u \\ \left(f - \text{sgn}(f)\dfrac{2A_z}{d_{amp}d_{rot}}\right)u = A_z\left(\dfrac{1}{d_{amp}^2} - \dfrac{1}{d_{rot}^2}\right)v + F_v \end{cases} \quad (S3)$$

In order to obtain non-trivial solutions for velocity and also a simplest but physically meaningful form for the additional force $F$, the following relations are necessary,

$$A_z = \frac{d_{amp}d_{rot}}{2}|f|, \ F_u = -Ru, \ F_v = -Rv, \ R = A_z\left(\frac{1}{d_{amp}^2} - \frac{1}{d_{rot}^2}\right) \quad (S4)$$

$F$ is in a form of Rayleigh friction that is often used in the analyses of atmospheric and oceanic dynamics although this force occurs everywhere in the settings herein, not just near a lateral or bottom boundary. Clearly, $R$ is a damping factor with dimension of $s^{-1}$. Instead of adding an additional force, $(F_u, F_v)$, in the Eqs. S2 above, McWilliams, Huckle [1] used a complex number for $A_z$. In the last section, we show that using a complex $A_z$ is equivalent to adding a linear damping term.

**Analytical solutions of the damping-enabled Ekman theory.**

The basic equations describing the Ekman layer dynamics are (equivalent to Eq. 1 of the main text)

$$\begin{cases} -fv = A_z \frac{\partial^2 u}{\partial z^2} - Ru \\ fu = A_z \frac{\partial^2 v}{\partial z^2} - Rv \end{cases} \quad (S5)$$

The correponsing boundary conditions are

$$\begin{cases} \rho_0 A_z \frac{\partial \mathbf{u}}{\partial z} = \boldsymbol{\tau}, \ z = 0 \\ \mathbf{u} = 0, \ z \to -\infty \end{cases} \quad (S6)$$

where $\rho_0 = $ constant is the reference density of seawater. Introducing the complex velocity $U = u + iv$, Eq. S5 becomes

$$ifU = A_z \frac{\partial^2 U}{\partial z^2} - RU$$

$$A_z \frac{\partial^2 U}{\partial z^2} - |f|(Da + i\,\text{sgn}(f))U = 0 \quad (S7)$$

where $Da = \dfrac{R}{|f|}$ is a non-dimensional number measuring the ratio between timescales of inertial motion and damping. The general solution for Eq. S5 is

$$U = U_s \exp(\lambda z) + U_{-s} \exp(-\lambda z) \quad (S8)$$

where $\lambda = \sqrt{\frac{|f|}{A_z}(Da + i\,\text{sgn}(f))}$ which can be written as $\lambda = \frac{\sqrt{2}}{\mu d_{Ek}}\exp(i\,\text{sgn}(f)\theta)$ with

$\mu = (1+Da^2)^{-\frac{1}{4}} = \sqrt{\sin 2\theta}$, $d_{Ek} = \sqrt{\frac{2A_z}{|f|}}$, and $\theta = \frac{1}{2}\arctan\left(\frac{1}{Da}\right)$. The boundary condition at

infinity (Eq. S6) requires $U_{-s}$ to be 0, then Eq. S8 becomes

$$U = U_s \exp\left[\frac{\sqrt{2}}{\mu d_{Ek}}(\cos\theta + i\,\text{sgn}(f)\sin\theta)z\right]$$
$$= U_s \exp\left(\frac{z}{d_{amp}} + i\,\text{sgn}(f)\frac{z}{d_{rot}}\right) \tag{S9}$$

where $d_{amp}$ and $d_{rot}$ are the depth scales for current amplitude and rotation

$$d_{amp} = \mu d_{Ek}\frac{\cos\frac{\pi}{4}}{\cos\theta}, \quad d_{rot} = \mu d_{Ek}\frac{\sin\frac{\pi}{4}}{\sin\theta} \tag{S10}$$

The product of the two depth scales is

$$d_{amp}d_{rot} = \frac{\mu^2}{2\sin\theta\cos\theta}d_{Ek}^2 = d_{Ek}^2 \tag{S11}$$

where $d_{Ek}$ is the depth scale in the classic Ekman theory. The boundary condition at the surface (Eq. S6) requires

$$\rho A_z \lambda U_s = |\tau|\exp(i\phi_\tau) \tag{S12}$$

where $|\tau|$ is the magnitude of wind stress, and $\phi_\tau$ is the direction of wind measured counterclockwise from the east. Substitution of $\lambda$ into Eq. S12 gives,

$$U_s = \frac{|\tau|}{\rho A_z}\frac{\mu d_{Ek}}{\sqrt{2}}\exp\left[i(\phi_\tau - \text{sgn}(f)\theta)\right]$$
$$= \mu U_{Ek}\exp\left[i(\phi_\tau - \text{sgn}(f)\theta)\right] \tag{S13}$$

where $U_{Ek} = d_{Ek}\frac{|\tau|}{\sqrt{2}\rho A_z}$ is the amplitude of surface current in the classic Ekman theory. $\theta$ in

Eq. S13 is the deflection angle of the surface current relative to the wind stress. Since $\theta = \frac{1}{2}\arctan\left(\frac{1}{Da}\right)$, it is always smaller than 45° for nonzero $R$. Eq. S13 also shows that the surface current is smaller than that predicted by the classic Ekman theory by a factor $\mu < 1$.

The Ekman transport with damping is the vertical integration of the velocity

$$T = T_x + iT_y = \int_{-\infty}^{0} U dz = \frac{U_s}{\lambda} = \mu^2 T_{Ek} \exp\left[i\left(\phi_\tau - \text{sgn}(f)2\theta\right)\right] \quad (S14)$$

where $T_{Ek} = \frac{|\tau|}{\rho|f|}$ is the magnitude of the classic wind driven Ekman transport. The Ekman pumping with damping is just the divergence of the transport $\frac{\partial T_x}{\partial x} + \frac{\partial T_y}{\partial y}$, ignoring the horizontal variation of $R$ gives

$$w = \mu^2\left[\nabla_z \cdot \left(\frac{\tau}{\rho|f|}\right)\cos 2\theta + \text{sgn}(f)\nabla_z \times \left(\frac{\tau}{\rho|f|}\right)\sin 2\theta\right]$$
$$= \mu^2\left[\nabla_z \cdot \left(\frac{\tau}{\rho|f|}\right)\cos 2\theta + w_{Ek}\sin 2\theta\right] \quad (S15)$$

where $w_{Ek} = \text{sgn}(f)\nabla_z \times \left(\frac{\tau}{\rho|f|}\right)$ is the Ekman pumping and the subscript $z$ in $\nabla_z$ means the operation is for the horizontal directions only. Hence, the Ekman pumping rate with damping includes the contribution from both the wind stress curl and wind stress divergence. When $R = 0$, $\theta = 45°$, $\mu = 1$, all solutions of the classic Ekman theory are recovered; $d_{amp} = d_{rot} = d_{Ek}$, there is only one depth scale.

**Inversing $d_{amp}$ and $d_{rot}$ from the observed Ekman spiral.**

The scale depth $d_{amp}$ determines the rate at which the current speed decays with depth. The current speed can be obtained from Eqs. S9 and S13,

$$|\mathbf{u}| = \sqrt{u^2 + v^2} = \mu U_{Ek} \exp\left(\frac{z}{d_{amp}}\right) \quad (S16)$$

Taking a natural logarithm of the equation above gives

$$\ln(|\mathbf{u}|) = \frac{1}{d_{amp}} z + \ln(\mu U_{Ek}) \qquad (S17)$$

Thus, a linear fit can be done to the observed current speed at different depths (Table S2), the slope of linear fit gives $\frac{1}{d_{amp}}$. Similarly, the scale depth $d_{rot}$ determines the rate at which the current direction rotates with depth. The current direction can be obtained from Eqs. S9 and S13 too,

$$\arctan\left(\frac{v}{u}\right) = \text{sgn}(f)\frac{1}{d_{rot}} z + \phi_\tau - \text{sgn}(f)\theta \qquad (S18)$$

$d_{rot}$ can also be obtained by a linear fit of the observed current direction at different depths. Given $d_{amp}$ and $d_{rot}$, the value of $A_z$ and $d_{Ek}$ can be obtained from Eq. S11, with which the value of $R$ can be obtained from Eq. S4. At last, the magnitude and angle of the wind stress can be obtained by substituting the observed Ekman current at a certain depth $[u_{obs}, v_{obs}, z_{obs}]$ into Eqs. S17 and S18.

The observed Ekman spirals from five locations are used (Table S1 and Fig. S2) in the inversion; their locations are at the south of the gulf stream[2], western Pacific near the equator[3], the coast of northern California[4], Drake Passage[5], and north of Kerguelen[6]. All the data are extracted from the figures in the relevant papers except those in [2], the original data of which were provided in [7]. The linear fits in Eqs. S17 and 18 for all five sites (Tables S1 and S2) are shown in the Fig. S1. Note that not all the data points at each observation site are used. For example, the observations at the south of the Gulf stream (site B in Fig. 2 of the main text) and equatorial Pacific (site C in Fig. 2 of the main text) were carried out early (before 1990), the vertical resolution was coarse and the uncertainty was probably large; the velocity at 15 m for site B and 20 m for site C have deflection angle larger than the velocity at a deeper depth. We chose to exclude these two data points from the observations. Fortunately, the wind stress at sites B and C was measured by mooring and ship-borne instruments which probably had a relatively high quality. For each of the other three observation sites, a data point at the top was removed to minimize the influence of surface waves and a data point at the bottom of the mixed layer was also excluded to minimize the influence of stratification.

**Model and configuration of the idealized experiment.**

The oceanic general circulation model MITgcm (https://mitgcm.readthedocs.io/en/latest/) is used to study the possible flattening of Ekman spiral. In our idealized experiment carried out herein, the advection is turned off in the momentum equation to be consistent with the Ekman model (Eq. S5). The experimental domain is rectangular with a dimension of 15.5 km × 15.5 km × 1260 m (Fig. S2). The grid spacing is uniformly 250 m in both directions; the spacing is 2 m for the first 25 layers and increased to 5 m for the next two layers, and then increased by 10 m consecutively until the bottom (layer 42) at which the layer thickness is 150 m. $f$-plane approximation is used with $f = 10^{-4}$ s$^{-1}$, equivalent to a latitude of 43°N. The water density is 1000 kg m$^{-3}$ and does not change with time and space. The vertical viscosity is constant and set to be $10^{-2}$ m$^2$ s$^{-1}$. The horizontal mixing coefficient $A_h$ is set to 400 m$^2$ s$^{-1}$. No-slip condition is specified at lateral boundaries; at the surface, zonal (i.e., along $x$ direction) wind stress is specified which has a value of 0.1 N m$^{-2}$ at y = 0 and decreases to 0 at the northern and southern boundaries according to a cosine function (Fig. S2).

To estimate $R$ in this case, we use $R = \frac{1}{2}\left(\frac{A_h \nabla^2 u}{u} + \frac{A_h \nabla^2 v}{v}\right)$ averaged over the first 10 m. This formula assumes that the damping of velocity is due to inhomogeneous horizontal shear. The value of $R$ is estimated to be $1.5 \times 10^{-5}$ s$^{-1}$. Damping may not be achieved this way in the ocean but it turns out this $R$ works very well here, as can be seen from the good fit between Ekman spirals predicted by the theory and simulated by the model (Fig. S3). The Ekman spiral shown in Fig. S3 is from the center of the simulation domain (Fig. S2).

**Model and configuration of the realistic global experiment.**

To see whether flattening of Ekman spirals in realistic simulations is comparable to that in the observations and whether the damping is indeed an important cause of the flattening, we utilize the results from a high-resolution simulation[8] with the Community Earth System model version 1.3 (CESM1.3). The simulation was done with a horizontal resolution of approximately 0.25° for the atmosphere and land and a nominal resolution of 0.1° for the ocean and sea-ice. The ocean component has 62 levels in the vertical with a maximum depth of 5875 m and the spacing is 10 m in the upper 100 m. The horizontal resolution is high enough so that the damping can be estimated using the second-order longitudinal structure function (see next section) and the vertical resolution is sufficient for demonstrating the Ekman spirals. The ocean wave model is stub only in the CESM1.3. The simulation is a "free-run" without any data assimilation and covers the period from 1850 to 2100, using time-varying historical greenhouse gas and anthropogenic and natural aerosol forcing from 1850 to 2005 and representative concentration pathway 8.5 (RCP8.5, a high greenhouse gas emission scenario) forcing during 2006–2100. More details on the model setting can be found in [8].

The simulation data nearest to the five locations of observations are analyzed and the Ekman spirals (purple curves in Fig. S4) are compared with the predictions from both the classic theory (blue curves in Fig. S4) and the new theory (orange curves in Fig. S4). For the location "Drake Passage", seven years of data are used, while for the other four locations, five years of data are used to improve the stability of spirals. Moreover, for the three locations that are far from the Southern Ocean, only the months when the direction of simulated winds is relatively stable are used; for the two locations in the Southern Ocean, the direction of wind is stable, and the data of the whole year are thus used. More details can be found in Table S3.

**Diagnosing damping in the realistic global experiment.**

It is essential to demonstrate that the damping in the model simulations is well linked to the damping ($R$ or $Da$) inferred from the simulated flatten Ekman spirals. However, because only monthly data are available from the realistic global simulation, the damping must be estimated indirectly from the kinetic energy dissipation rate. For this purpose, the second-order longitudinal structure function (see Eq. S19[9]) is a powerful tool,

$$D_{LL}(r) = \left\langle \left\{ [u(x+r) - u(x)] \cdot \hat{r} \right\}^2 \right\rangle = C_2 \varepsilon^{\frac{2}{3}} r^{\frac{2}{3}} \tag{S19}$$

Here $D_{LL}$ is the second-order longitudinal structure function. $[u(x+r) - u(x)] \cdot \hat{r}$ is the longitudinal velocity differences where $r$ is the vector connecting the two points. $C_2$ is Kolmogorov constant; $\varepsilon$ is the energy dissipation rate. Kolmogorov[10] predicted that the $D_{LL}$ obeys the "two-thirds law" when the homogeneous and isotropic turbulence fluid is in the inertial range where the effect of intermittency can be negligible. Although $C_2$ is called a constant, it varies with the Reynolds number and ranges from 1.1 to 2.5[9]. This might induce some uncertainty to the estimated energy dissipation rate. To minimum this uncertainty, we combine $C_2$ and the energy dissipation rate $\varepsilon$ to define a new index $\xi$ to reflect the degree of dissipation. Moreover, the vector $r$ in the model can be presented with $r = n\Delta$ where $n$ is the grid point counted from the center $x$ and $\Delta$ is the longitudinal spacing (~10 km). Therefore, the new index is defined as

$$\xi = D_{LL}/n^{\frac{2}{3}} = C_2 \varepsilon^{\frac{2}{3}} \Delta^{\frac{2}{3}} \tag{S20}$$

Fig. S5 shows the variation $\xi$ with $n$ (equivalent to $r$), in which $\xi$ is approximately constant when $n$ is large (in the inertial range[9]). This feature confirms the appropriateness of using $\xi$ to infer damping, although it cannot give the value of $\varepsilon$ directly. The shape of the curve for the south of Gulf stream is abnormal, with the value of $\xi$ varying violently when $n$ is large (Fig. S5B). This may be because of the strong anisotropy in the flow field, violating the assumption made in obtaining Eq. (S19).

**Equivalence between lateral friction and complex vertical viscosity.**

For a complex viscosity, the Ekman equation has its classic form,

$$ifU = A_z \frac{\partial^2 U}{\partial z^2} \tag{S21}$$

where $A_z = A_{zr} + iA_{zi}$ is the complex vertical viscosity (similar to Eq. B1 of McWilliams, Huckle [1]).

The general solution for this equation is

$$U = U_s \exp(\lambda z), \lambda^2 = i\frac{f}{A} = i\frac{f}{A_{zr}^2 + A_{zi}^2}(A_{zr} - iA_{zi}) \tag{S22}$$

So that $\frac{\partial^2 U}{\partial z^2} = \lambda^2 U = i\frac{f}{A_{zr}^2 + A_{zi}^2}(A_{zr} - iA_{zi})U$ and Eq. S21 becomes

$$ifU = A_{zr}\frac{\partial^2 U}{\partial z^2} + iA_{zi}\frac{\partial^2 U}{\partial z^2}$$

$$ifU = A_{zr}\frac{\partial^2 U}{\partial z^2} - \frac{f}{A_{zr}^2 + A_{zi}^2}\left(A_{zr}A_{zi} - iA_{zi}^2\right)U$$

$$if\frac{A_{zr}^2}{A_{zr}^2 + A_{zi}^2}U = A_{zr}\frac{\partial^2 U}{\partial z^2} - f\frac{A_{zr}A_{zi}}{A_{zr}^2 + A_{zi}^2}U \qquad (S23)$$

$$ifU = \frac{A_{zr}^2 + A_{zi}^2}{A_{zr}}\frac{\partial^2 U}{\partial z^2} - f\frac{A_{zi}}{A_{zr}}U$$

$$ifU = A_{ze}\frac{\partial^2 U}{\partial z^2} - RU$$

where $A_{ze} = A_{zr}\left(1 + \frac{A_{zi}^2}{A_{zr}^2}\right)$ and $R = f\frac{A_{zi}}{A_{zr}}$. Eq. S23 now has the same format as Eq. 1 of the main text. Both $A_{ze}$ and $R$ are real numbers with $R$ equivalent to the damping factor in the main text.

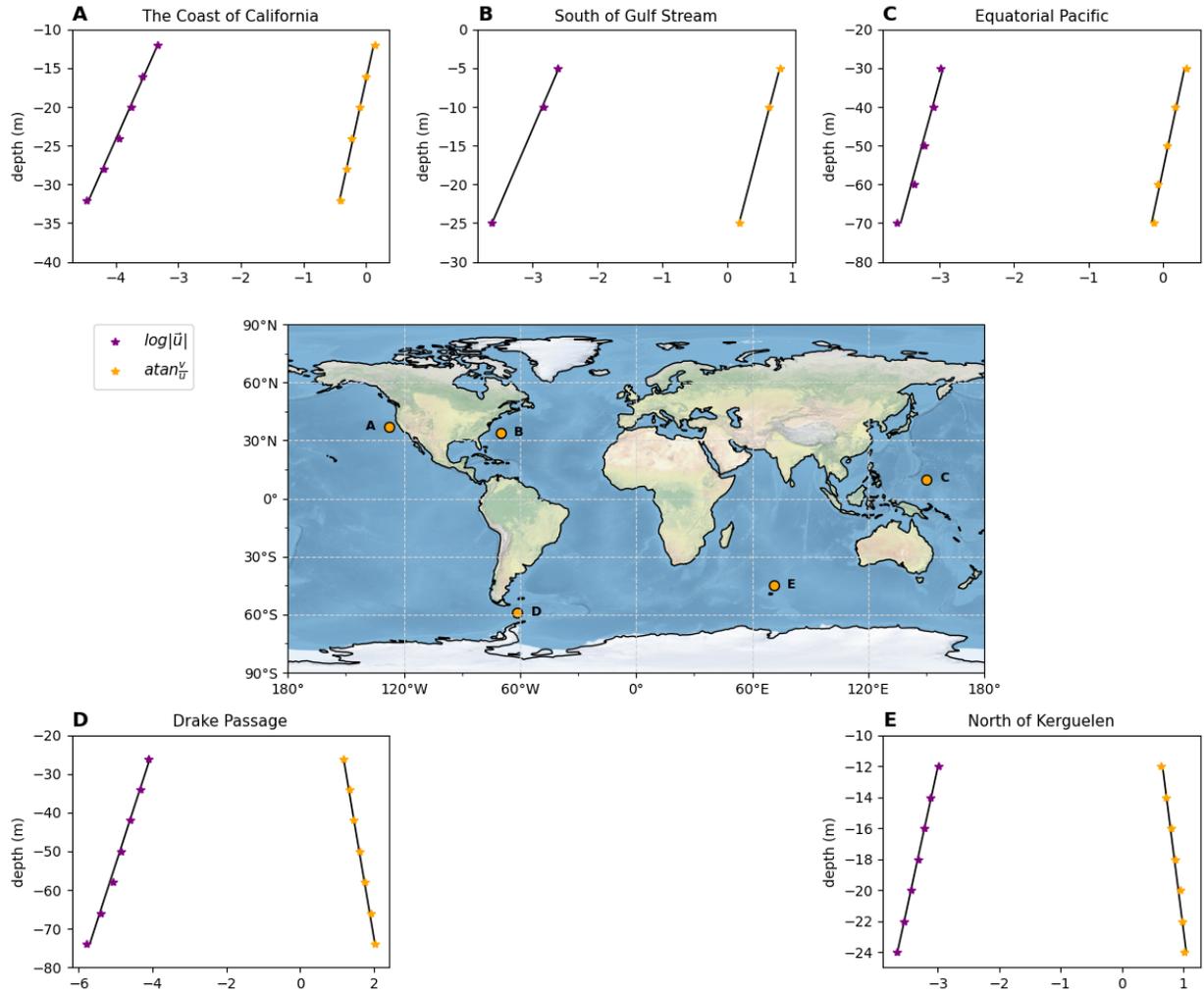

**Fig. S1.** Linear fitting between (purple stars) the magnitude ($ln(|u|)$ in Eq. S17) and (orange stars) the direction ($\arctan\left(\dfrac{v}{u}\right)$ in Eq. S18) and the depth of Ekman currents from the five observation sites in Tables S1 and S2, and Fig. 2 of the main text.

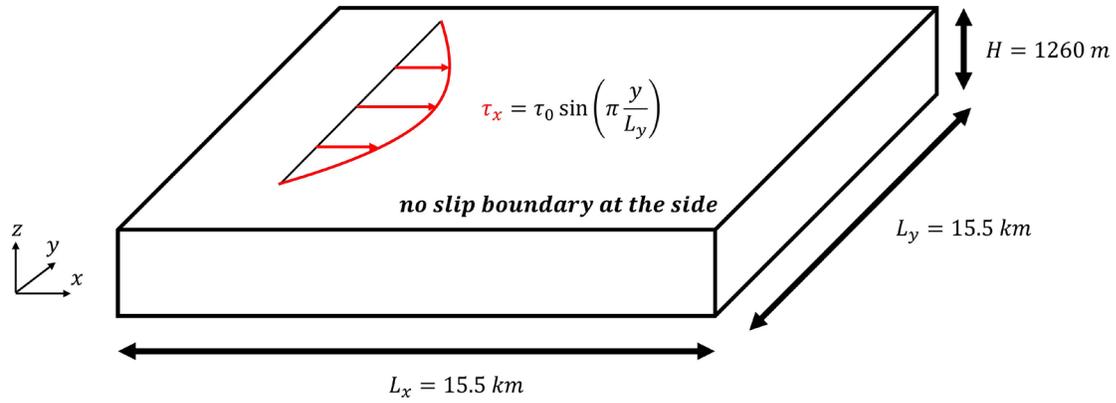

**Fig. S2.** The domain and configuration of the MITgcm experiment. The wind stress applied at the surface is also indicated on the left.

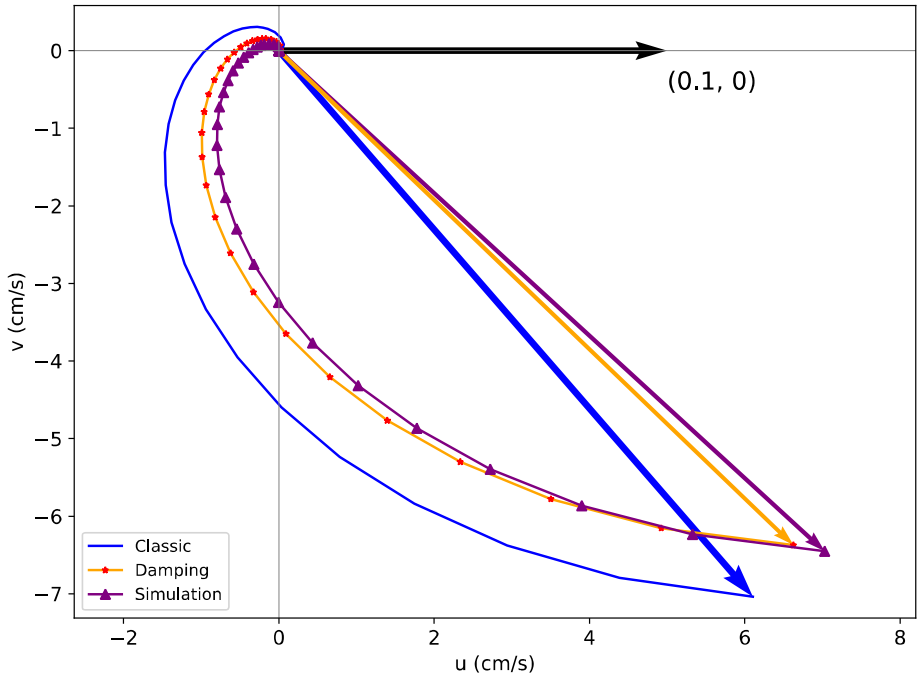

**Fig. S3.** Comparison between the simulated Ekman spirals (purple triangles; from experiment shown in Fig. S2) and the theoretical ones (blue: classic Ekman theory; orange: damping-enabled theory). The black arrow depicts the wind stress in unit of N/m$^2$; the vertical spacing of data points is 2 m and the first point is at 2 m depth.

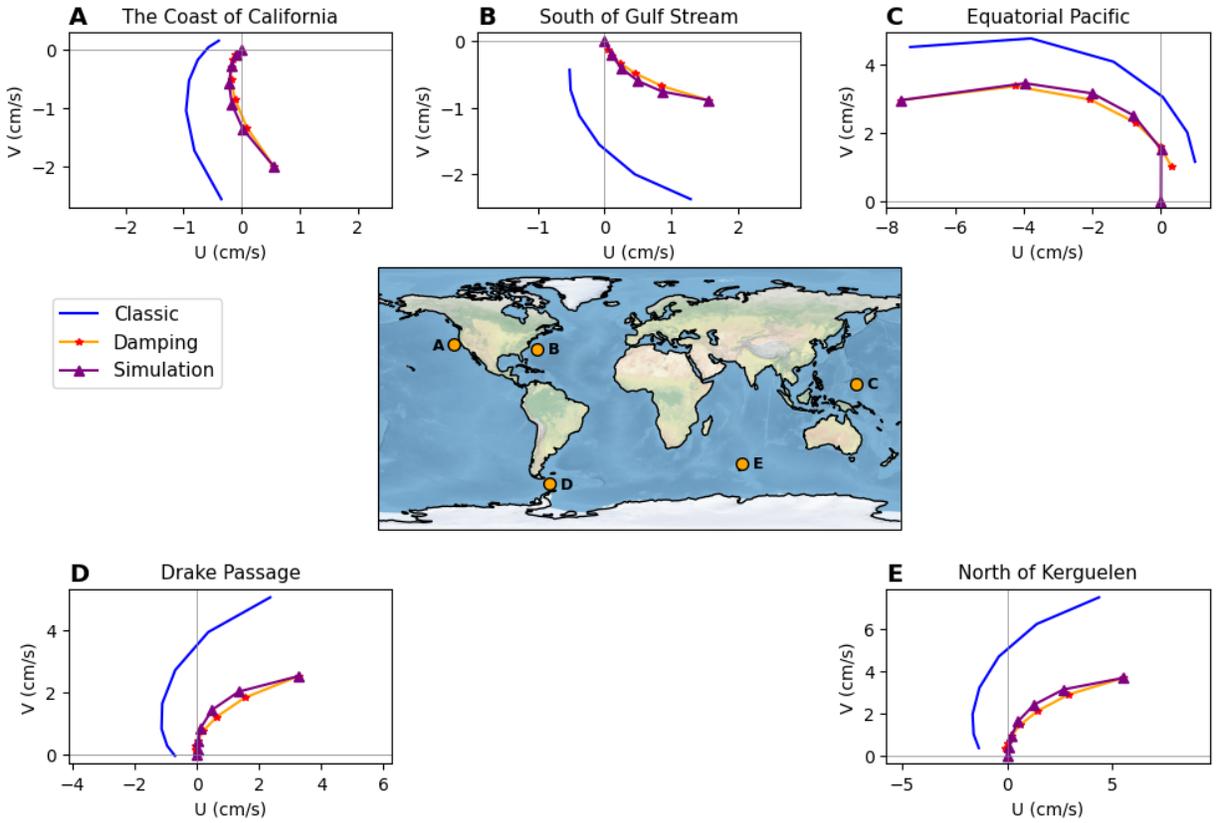

**Fig. S4. Comparison between the simulated Ekman spirals (purple triangles) in CESM1.3 and the theoretical ones (blue: classic Ekman theory; orange: damping-enabled theory).** The velocities at depths from 5 m to 55 m (b and c) or 65 m (a, d, and e) with an interval of 10 m are shown. The x-axis and y-axis are along east and north, respectively, in all panels, different from those in the Fig. 2 of the main text.

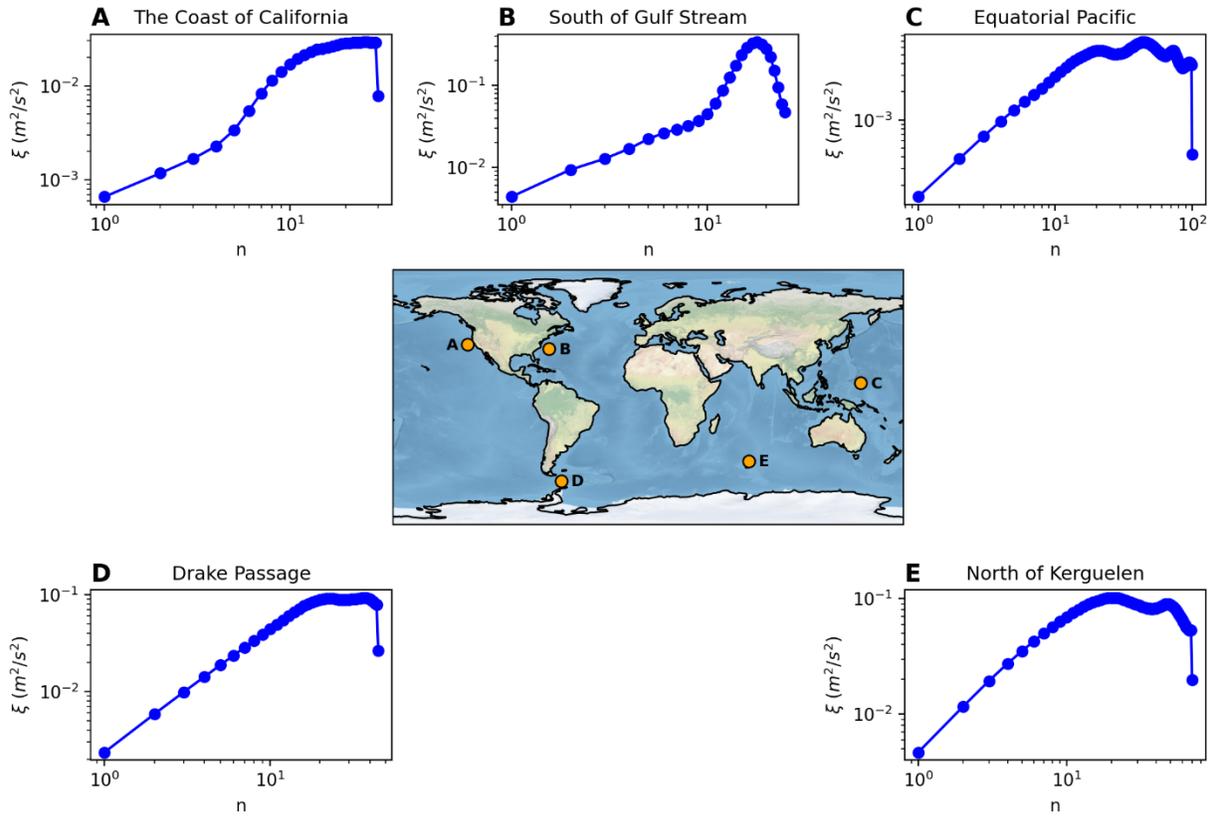

**Fig. S5. The dissipation rate calculated at five sites where the observations of Ekman spirals were made (Fig. 2 of the main text).** The calculation method can be found in Eq. S20. The value of $\xi$ when the curves become flat is used as the dissipation rate.

**Table S1.** The observed Ekman currents. The grey shadow area represents that the data points that were used to inverse $d_{amp}$ and $d_{rot}$ (Fig. S1 and Table S2). "*" represents the data point that was used to calculate wind stress. All these data points are shown in Fig. 2 of main text.

| The Coast of California | Depth(m) | -8 | -12* | -16 | -20 | -24 | -28 | -32 | | |
|---|---|---|---|---|---|---|---|---|---|---|
| | U(cm/s) | -4.43 | -3.51 | -2.79 | -2.33 | -1.87 | -1.44 | -1.05 | | |
| | V(cm/s) | -1.45 | -0.51 | -0.01 | 0.21 | 0.42 | 0.47 | 0.44 | | |
| South of the Gulf Stream | Depth(m) | -5* | -10 | -25 | -50 | -75 | | | | |
| | U(cm/s) | 5.0 | 4.7 | 2.6 | 2.1 | 0.7 | | | | |
| | V(cm/s) | 5.4 | 3.5 | 0.5 | -0.1 | -0.3 | | | | |
| Equatorial Pacific | Depth(m) | -30* | -40 | -50 | -60 | -70 | -80 | | | |
| | U(cm/s) | 4.79 | 4.5 | 4.01 | 3.51 | 2.79 | 1.3 | | | |
| | V(cm/s) | 1.57 | 0.79 | 0.26 | -0.21 | -0.34 | -0.24 | | | |
| Drake Passage | Depth(m) | -26* | -34 | -42 | -50 | -58 | -66 | -74 | -82 | -90 |
| | U(cm/s) | 0.62 | 0.32 | 0.13 | -0.02 | -0.1 | -0.15 | -0.13 | -0.1 | -0.05 |
| | V(cm/s) | 1.5 | 1.26 | 1.00 | 0.78 | 0.61 | 0.42 | 0.27 | 0.14 | 0.05 |
| North of Kerguelen | Depth(m) | -10 | -12* | -14 | -16 | -18 | -20 | -22 | -24 | |
| | U(cm/s) | 4.8 | 4.04 | 3.32 | 2.78 | 2.32 | 1.87 | 1.58 | 1.36 | |
| | V(cm/s) | 2.63 | 3.01 | 2.94 | 2.89 | 2.76 | 2.62 | 2.42 | 2.19 | |

**Table S2.** Calculated and observed variables and parameters at five sites where Ekman spirals were observed.

| Location | The coast of California | South of the Gulf Stream | Equatorial Pacific | Drake Passage | North of Kerguelen |
|---|---|---|---|---|---|
| (Lat, Lon) | (37.1°N, 127.6°W) | (34°N, 70°W) | (10°N, 140°E-160°E) | (54°S-64°S, 66°W -57°W) | (41°S-49°S, 64°E-78°E) |
| Date (mm/yy) | 04/93-10/93 | 11/82-02/83 | 02/89-03/89 | 09/99-10/06 | 11/08-01/09 |
| $d_{amp}$ (m) | 18.2±0.7 | 19.4±0.7 | 70.0±6.5 | 29.4±1.3 | 18.2±0.2 |
| $d_{rot}$ (m) | 37.0±1.5 | 32.0±1.4 | 90.0±6.4 | 56.0±1.2 | 31.0±2.3 |
| $A_z$ ($10^{-2}$ m$^2$ s$^{-1}$) | 2.9±0.2 | 2.5±0.1 | 8.0±0.9 | 10.4±0.5 | 2.9±0.2 |
| $R$ ($10^{-5}$ s$^{-1}$) | 6.7±0.7 | 4.2±0.5 | 0.7±0.3 | 8.8±0.7 | 5.8±0.9 |
| $f$ ($10^{-4}$ s$^{-1}$) | 0.877 | 0.813 | 0.253 | -1.26 | -1.03 |
| Da | 0.76 | 0.52 | 0.28 | 0.70 | 0.56 |
| $N$ ($10^{-3}$ s$^{-1}$) | 5.23 | 1.64 | 3.03 | 1.46 | 0.99 |
| $H$ (m) | 20.8 | 113.7 | 82.7 | 92.8 | 151.0 |
| EKE (cm$^2$ s$^{-2}$) | 155±29 | 212±47 | 93±87 | 201±34 | 159±48 |
| $EKE_{obs}$ (cm$^2$ s$^{-2}$) | 189.53 | 397 | 117 | 206 | 394 |
| $\tau_{mag}$ (N m$^{-2}$) | 0.127 | 0.148 | 0.114 | 0.161 | 0.185 |
| $\tau_{obs}$ (N m$^{-2}$) | 0.09 | 0.147 | 0.11 | 0.08 | 0.099 |

**Table S3.** The damping estimated from the global simulation data for the same locations as in Table S1 where Ekman spirals were observed. $R$ and $Da$ are estimated by inversion from the simulated Ekman spirals, while $\xi$ is estimated utilizing the turbulence theory from the simulated surface velocity (Fig. S5). The first row also gives the time periods of the simulations used to calculate these parameters as well as the Ekman spirals (Fig. S4).

| Location | The coast of California | South of the Gulf Stream | Equatorial Pacific | Drake Passage | North of Kerguelen |
|---|---|---|---|---|---|
| Year | 1991-1995 | 1981-1985 | 1987-1991 | 1999-2006 | 2007-2011 |
| month | 2-4 | 1,11,12 | 2-3 | 1-12 | 1-12 |
| $R$ ($10^{-5}$ s$^{-1}$) | 7.62 | 12.2 | 0.81 | 13.7 | 10.6 |
| $Da$ | 0.869 | 1.502 | 0.32 | 1.087 | 1.031 |
| $\xi$ ($10^{-2}$ m$^2$s$^{-2}$) | 2.88 | 25.4 | 0.608 | 8.93 | 8.47 |

# References

1. McWilliams JC, Huckle E, Liang J-H *et al.* The wavy Ekman layer: Langmuir circulations, breaking waves, and Reynolds stress. *Journal of Physical Oceanography*. 2012; **42**(11): 1793-1816.
2. Price JF, Weller RA, Schudlich RR. Wind-driven ocean currents and Ekman transport. *Science*. 1987; **238**(4833): 1534-1538.
3. Wijffels S, Firing E, Bryden H. Direct observations of the Ekman balance at 10 N in the Pacific. *Journal of Physical Oceanography*. 1994; **24**(7): 1666-1679.
4. Chereskin T. Direct evidence for an Ekman balance in the California Current. *Journal of Geophysical Research: Oceans*. 1995; **100**(C9): 18261-18269.
5. Lenn Y-D, Chereskin TK. Observations of Ekman currents in the Southern Ocean. *Journal of Physical Oceanography*. 2009; **39**(3): 768-779.
6. Roach CJ, Phillips HE, Bindoff NL *et al.* Detecting and characterizing Ekman currents in the Southern Ocean. *Journal of Physical Oceanography*. 2015; **45**(5): 1205-1223.
7. Schudlich RR, Price JF. Observations of seasonal variation in the Ekman layer. *Journal of physical oceanography*. 1998; **28**(6): 1187-1204.
8. Chang P, Zhang S, Danabasoglu G *et al.* An unprecedented set of high-resolution earth system simulations for understanding multiscale interactions in climate variability and change. *Journal of Advances in Modeling Earth Systems*. 2020; **12**(12): e2020MS002298.
9. Ni R, Xia K-Q. Kolmogorov constants for the second-order structure function and the energy spectrum. *Physical Review E—Statistical, Nonlinear, and Soft Matter Physics*. 2013; **87**(2): 023002.
10. Kolmogorov AN. The local structure of turbulence in incompressible viscous fluid for very large Reynolds numbers. *Proceedings of the Royal Society of London Series A: Mathematical and Physical Sciences*. 1991; **434**(1890): 9-13.